# Dynamic mitigation of filamentation instability and magnetic reconnection in sheet-current sustained plasma

[1, 2] Yan-Jun Gu, [3] Shigeo Kawata and [2, 4] Sergei V. Bulanov

[1] Institute of Laser Engineering, Osaka University, Suita, Osaka 565-0871, Japan
[2] Institute of Physics of the ASCR, ELI-Beamlines, Na Slovance 2, 18221 Prague, Czech Republic.
[3] Graduate School of Engineering & Graduate School of Regional Development and Creativity, Utsunomiya University, Yohtoh 7-1-2, Utsunomiya 321-8585, Japan.
[4] Kansai Photon Science Institute, National Institutes for Quantum and Radiological Science and Technology, 8-1-7 Umemidai, Kizugawa-shi, Kyoto 619-0215, Japan.

Abstract

Dynamic mitigation is presented for filamentation instability and magnetic reconnection in a plasm driven by a wobbling electron sheet current. The wobbling current introduces an oscillating perturbation and smooths the perturbation. The sheet current creates an anti-parallel magnetic field in plasma. The initial small perturbation induces the electron beam filamentation and the magnetic reconnection. When the wobbling or oscillation motion is added to the sheet electron beam along the sheet current surface, the perturbation phase is mixed and consequently the instability growth is delayed remarkably. Normally plasma instabilities are discussed by the growth rate, because it would be difficult to measure or detect the phase of the perturbations in plasmas. However, the phase of perturbation can be controlled externally, for example, by the driver wobbling motion. The superimposition of perturbations introduced actively results in the perturbation smoothing, and the instability growth can be reduced, like feed-forward control.

Dynamic mitigation is presented for sheet-current driven plasma instability. Theory and 3-dimensional (3-D) simulations show a clear mitigation of the electron current filamentation instability growth [1-5] and the magnetic reconnection [6-16]. Plasma instabilities emerge from perturbations. Superimposition of phase-controlled plasma perturbations can reduce the instability growth, like feed-forward control [17, 18]. The dynamic mitigation of plasma and fluid instability was proposed in Refs. [19-22]. An energy-carrying driver would introduce perturbations into plasma systems. If the perturbation phase is controlled by, for example, wobbling motion of driver beam, the superimposed overall perturbation amplitude can be reduced.

A sheet electron beam current in plasma creates an anti-parallel magnetic field with a magnetic shear along the electron current sheet as shown in Fig. 1(a). The sheet electron beam would cause filamentation instability and associated magnetic reconnection. Sheet-electron sustained magnetic shear is found in various situations: for example, it exists at solar flare, near-earth space, and magnetic confinement fusion. The magnetic reconnection would create energetic charged particles.

The equilibrium state for the sheet current plasma is based on the Harris solution in Ref. [23]. In our system, the electron averaged current is directed to -$x$ and the electron sheet is located in the $x - y$ plane. The density peak is at $z=0$ (see Fig. 1(b)). The averaged number density $n$ is obtained by $n = n_0/\cosh^2(\beta z/\lambda_D)$. Here $-2\beta c$ shows the averaged electron speed in $x$, $c$ the speed of light, $\beta = v/c$, and $\lambda_D$ the Debye length. In this work the initial temperature is 10eV for electrons and ions, and the plasma is fully ionized hydrogen. The protons are stationary at $t = 0$, and the initial $\beta$ is 0.1. Initially the electron number density $n_e$ is set to $n_e = n_i + 2\gamma\beta^2 n$, where the second term of the righthand side comes from the Lorentz transformation. The initial magnetic field $\vec{B}$ is obtained by $\vec{B} = B_0 \times (0, \tanh(\beta z/\lambda_D), 0)$ (see Fig. 1(c)), and the electric field is expressed by $\vec{E} = (0, 0, \gamma\beta B_y)$ at $t = 0$. The initial small perturbation is imposed in $n_e$ along $y$ with the amplitude of 5% and the wavelength of $L_y = 0.1$ m. In our simulations we employ 3D particle-in-cell EPOCH [23]. The simulation box is 0.2m×0.2m×0.2m, and $n_0$ is $1.0 \times 10^{14}$ m$^{-3}$. The corresponding mesh cells are 200×200×200 with 64 particles in each cell. In $x$ and $y$, the periodic boundary conditions are employed, with the ideal conducting boundary conditions in $z$. The scale lengths of $\lambda_D/\beta$ and $L_y$ in our system are smaller than the electron inertial length $c/\omega_{pe}$.

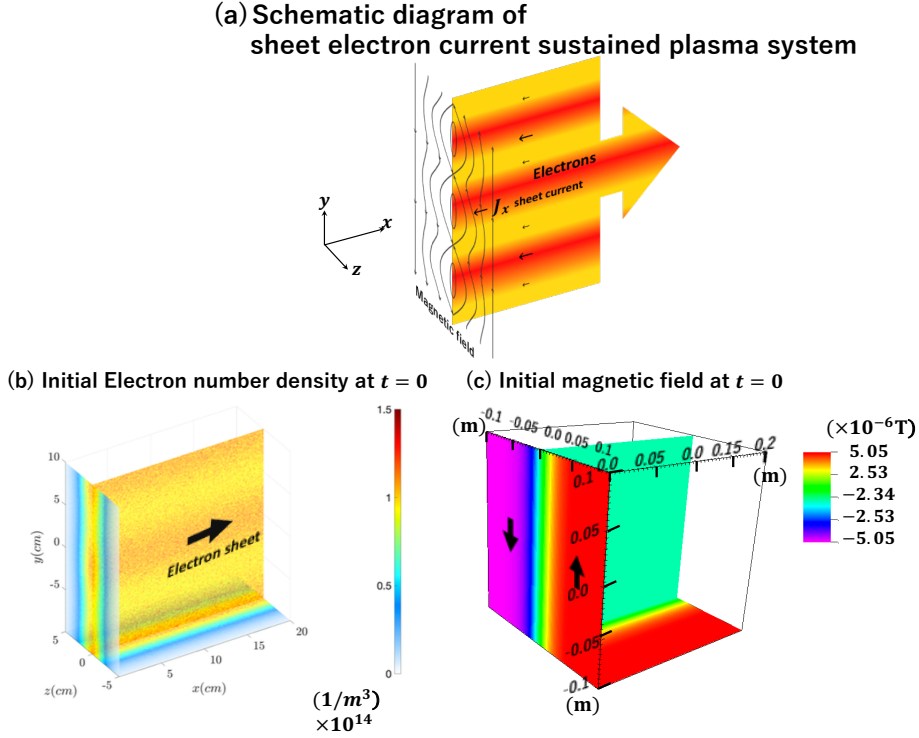

FIG. 1. (a) Schematic diagram of plasma system sustained by sheet electron current. The magnetic reconnection is induced along with the plasma filamentation. The initial conditions of (b) the electron number density $n_e$ and (c) the magnetic field are also presented. The electron sheet current is located near $z=0$ and the electrons move in the $+x$ direction. A small electron density perturbation is imposed in $n_e$ along $y$.

Figures 2 show the spatial distributions of the electron number density $n_e$ at (a) $t = 0.5$μs and (b) $t = 0.7$μs, and of the proton density $n_i$ at (c) $t = 0.0$μs, (d) $t = 0.5$μs and (e) $t = 0.7$μs. The electron sheet current is filamented along the $z=0$ plane, and the ions also follow the filamentation. Figures 3 present the distributions of magnetic field strength at $x$=0 at (a)$t = 0.0$μs, (b)$t = 0.5$μs, (c)$t = 0.7$μs and (d)$t = 0.8$μs. At $z$=0 the initial magnetic shear appears. The filamentation and the magnetic reconnection become remarkable around $t = 0.5$μs. The growth rate of the magnetic reconnection may be estimated by $\gamma_{rec} \sim v_A/L$ [14], which is about $\gamma_{rec} \sim 2.18 \times 10^6$/s in our cases. The Alfven speed $v_A$ is $v_A \sim 2.18 \times 10^4$ m/s, and the scale length $L$ would be about 0.01m. The growth time scale would be estimated by $\tau \sim 1/\gamma_{rec} \sim 0.458$μs. Around $t = 0.8$μs it may enter the nonlinear stage. Figures 4 show the magnetic field vectors at $x$=0 at (a) $t = 0.6$μs and (b)$t = 0.75$μs. In Figs. 5, the net current distributions are shown near the electron current $(J_x)$ sheet at (a)$t = 0.5$μs and (c)$t = 1.0$μs. The current density $J_x$ is normalized by $n_0 ec$. The net current shows the filamentation along with the magnetic reconnection.

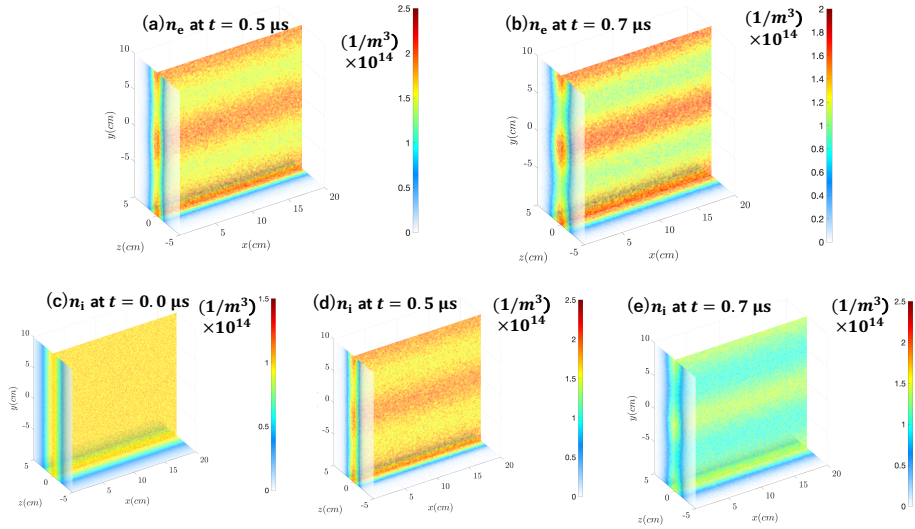

FIG. 2. Spatial distributions of the electron number density $n_e$ at (a)$t = 0.5$μs and (b) $t = 0.7$μs, and the proton density $n_i$ at (c)$t = 0.0$μs, (d)$t = 0.5$μs and (e) $t = 0.7$μs. The electron sheet current is filamented along the $z$ =0 plane, and the ions also follow the filamentation.

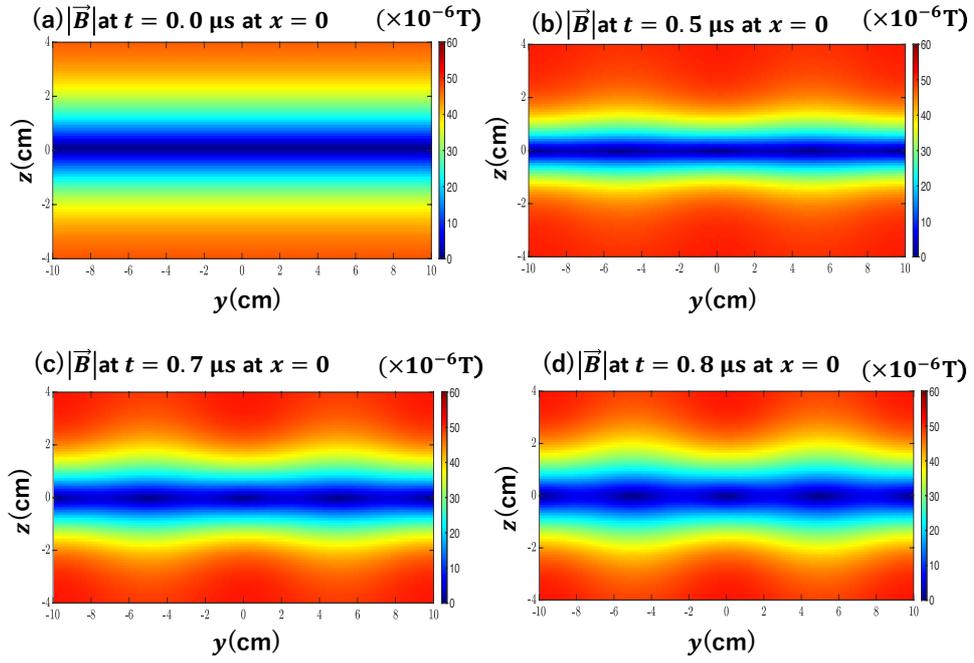

FIG. 3. Distributions of magnetic field strength at $x=0$ at (a) $t = 0.0$μs, (b) $t = 0.5$μs, (c) $t = 0.7$μs and (d) $t = 0.8$μs. Around $t = 0.5\sim 0.7$μs the magnetic reconnection becomes distinct.

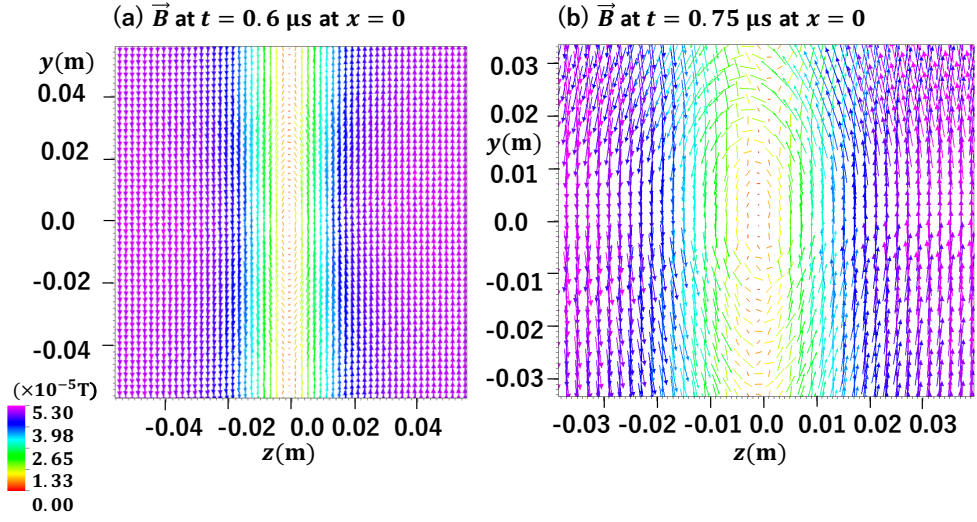

FIG. 4. Distributions of magnetic field vectors at $x=0$ at (a) $t = 0.6$μs and (b) $t = 0.75$μs. At $t = \sim 0.5 \sim 0.75$μs the magnetic reconnection becomes distinct.

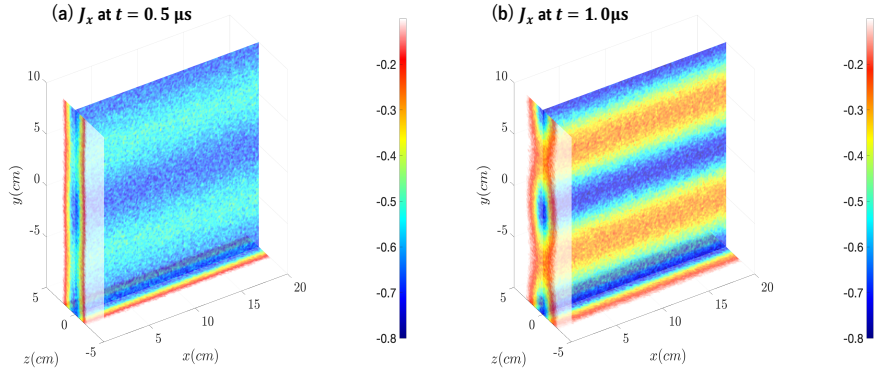

FIG. 5. Distributions of current near the electron current sheet $J_x$ at (a) $t = 0.5$μs and (b) $t = 1.0$μs. The net current shows the filamentation along with the magnetic reconnection.

The dynamic mitigation mechanism [19-22] is applied to the sheet-current driven magnetized plasma. The dynamic mitigation mechanism contributes to mitigate instabilities and also to smooth the non-uniformity of physical quantities in plasmas and fluids. When a physical quantity $\phi$ is perturbed like $\delta\phi = \delta\phi_0 e^{\gamma(t-\tau)+i\vec{k}\cdot\vec{x}+i\Omega\tau}$ in an unstable system, the perturbation of $\delta\phi$ grows with the growth rate of $\gamma > 0$. Here $\delta\phi_0$ is the perturbation amplitude, $\vec{k}$ the wave number vector, $\tau$ the time at which the perturbation is applied, and $\Omega$ defines the perturbation phase. In plasmas, it would be difficult to measure the perturbation phase, and therefore, the feedback control [17] cannot be directly applied to control plasma instability growth. However, the perturbation phase $\Omega\tau$ would be defined externally by, for example, energy-carrying driver oscillation. When the perturbations introduced at $t = \tau$ change the phase continuously by $\Omega\tau$, the overall perturbation superimposed at $t$ is obtained by $\int_0^t d\tau\ \delta\phi_0 e^{i\Omega\tau} e^{\gamma(t-\tau)+i\vec{k}\cdot\vec{x}} \propto \frac{\gamma+i\Omega}{\gamma^2+\Omega^2}\delta\phi_0 e^{\gamma t} e^{i\vec{k}\cdot\vec{x}}$. Although the growth rate $\gamma$ does not change, the perturbation amplitude is well reduced by the factor of $\sim\gamma/\Omega$ for $\Omega > \gamma$, compared with that without the phase oscillation. The theoretical consideration suggests that the frequency $\Omega$ in the perturbation phase change should be larger than or at least comparable to $\gamma$ for the effective mitigation or smoothing of the perturbations.

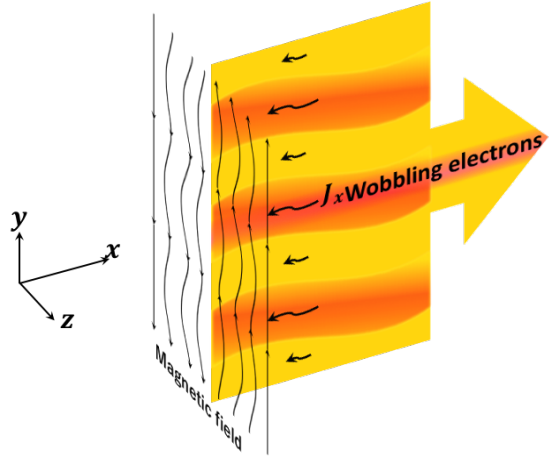

FIG. 6. Schematic diagram for dynamic phase control in electron current sheet sustained plasma system, in which electron filamentation and magnetic reconnection grow. The electron sheet current is oscillated along the sheet, and the perturbation phases introduced by the electron beam smooth and mitigate the perturbation amplitude.

The phase-control dynamic mitigation mechanism is applied to the electron sheet current plasma system in Fig. 1. The sheet electron beam oscillates along the sheet along $y$ with the amplitude of 0.1m, the wavelength in $x$ of 0.1m and the wobbling frequency $\Omega$ of 300MHz, which is large compared with the reconnection growth rate $\gamma_{rec}$: $\Omega \gg \gamma_{rec}$. Figure 6 shows the schematic diagram for the dynamic phase control in electron current sheet sustained plasma system in Fig. 1. The electron sheet current is oscillated along the sheet, and it is expected that the perturbation phases introduced by the electron beam smooth and mitigate the perturbation amplitude.

Figures 7 show the electron number density $n_e$ at (a)$t = 0.0\mu s$, (b) $t = 0.5\mu s$ and (c)$t = 0.7\mu s$, and the proton density $n_i$ at (d)$t = 0.5\mu s$ and (e) $t = 0.7\mu s$. Figure 7(a) presents the oscillation or wobbling motion of the electron sheet current. In Figs. 7 the filamentation does not grow well compared with the results in Figs. 2. Figures 8 show the magnetic field strength at $x=0$, and the magnetic field reconnection is not remarkable. Comparing with the results in Figs. 3, the dynamic mitigation or onset delay of the magnetic reconnection is clearly shown in Figs. 8. Figures 9 show the net current distributions $J_x$ at (a)$t = 0.5\mu s$ and (b)$t = 1.0\mu s$. By the wobbling or oscillating motion of the sheet electron current along $y$, the onsets of the filamentation and the magnetic reconnection are delayed and mitigated clearly.

In order to compare the filamentation and magnetic reconnection in the electron sheet current sustained plasma system schematically shown in Figs. 1(a) and 6, the histories of the normalized field energy of $B_z^2$ are presented in Fig. 10. In our systems in Figs. 1(a) and 6, the major energy is carried by the electron kinetic energy, and it creates the magnetic field shown in Fig. 1(b). The electron current is filamented along with the magnetic reconnection. Associated with the electron current filamentation and the magnetic reconnection, $B_z$ is induced. Figure 10 presents the clear difference in the field energy between the two cases.

With the wobbling motion of the electron beam (see Figs. 6 and 7(a)), Fig. 10 demonstrates the onset delay of the filamentation and magnetic reconnection in the sheet current sustained plasma system.

The theoretical considerations and 3D numerical computations show the clear effectiveness and viability of the dynamic phase control method to mitigate the plasma instability and the magnetic reconnection. The sheet current plasma system can be found in magnetic fusion devices, space, terrestrial magnetic system, etc. The dynamic mitigation mechanism may contribute to mitigate the magnetic fusion plasma disruptive behavior or to understand the stable structure of the sheet current sustained plasma system. In this paper we focus on the non-relativistic magnetic reconnection and the filamentation (tearing mode like) instability, and the major energy carrier is the sheet electron current. On the other hand relativistic magnetic reconnection has been also studied in space, solar system, planetary magnetic field, etc.[7, 26, 27] In those cases, the main energy is carried by the magnetic field and the electromagnetic field. In the relativistic

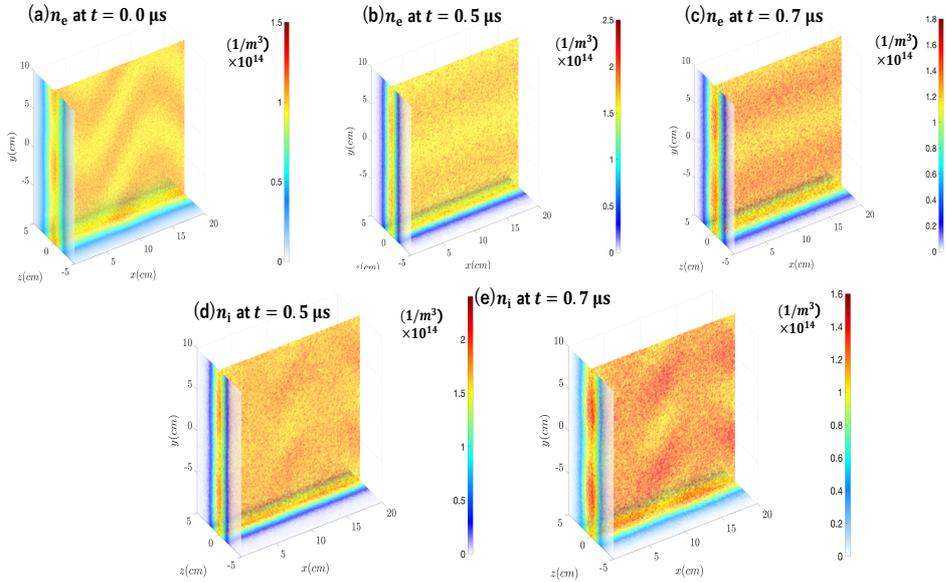

FIG. 7. Electron number density $n_e$ at (a) $t = 0.0$μs, (b) $t = 0.5$μs and (c) $t = 0.7$μs, and the proton density $n_i$ at (d) $t = 0.5$μs and (e) $t = 0.7$μs. Figure 7(a) presents the oscillation or wobbling motion of the electron sheet current. In Figs. 7 the filamentation does not grow well compared with the results in Figs. 2.

magnetic reconnection case, the dynamic mitigation mechanism should be further studied.

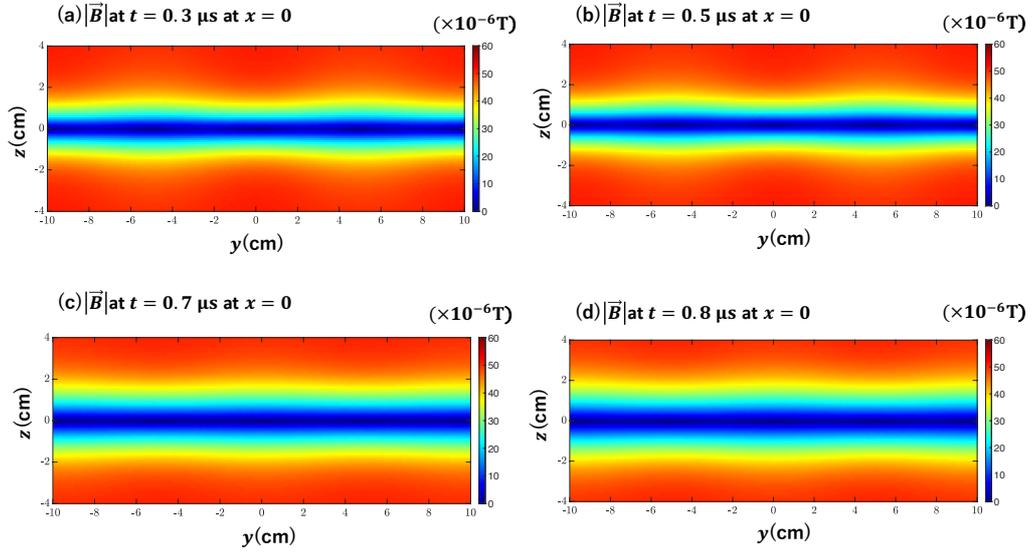

FIG. 8. Distributions of magnetic field strength at $x=0$ at (a)$t = 0.3$μs, (b)$t = 0.5$μs, (c)$t = 0.7$μs and (d)$t = 0.8$μs. By the electron wobbling motion along $y$, the magnetic reconnection is mitigated well.

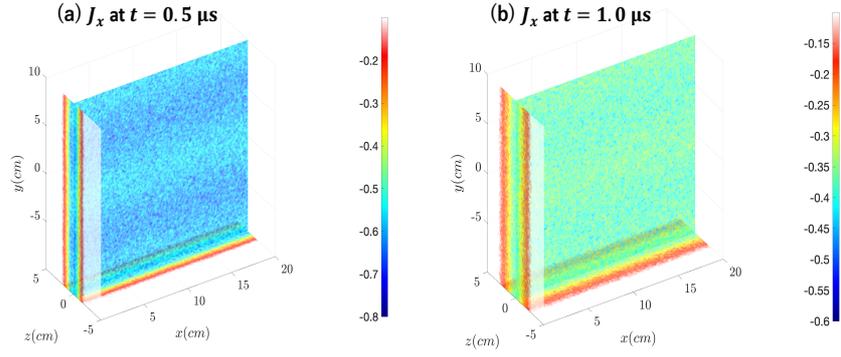

FIG. 9. Distributions of current $J_x$ near the electron current sheet at (a)$t = 0.5$μs and (b)$t = 1.0$μs. By the wobbling or oscillating motion of the sheet electron current along $y$, the onsets of the filamentation and the magnetic reconnection are delayed and mitigated clearly.

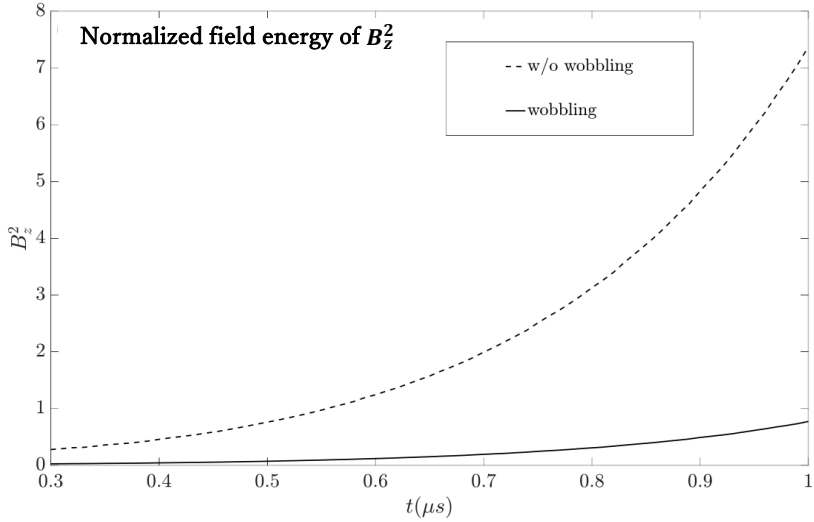

FIG. 10. Histories of normalized field energy of $B_z^2$ for the sheet electron current plasma systems with (solid line) and without (dotted line) the wobbling or oscillating motion of the sheet electron current along $y$. The onsets of the filamentation and the magnetic reconnection are delayed and mitigated clearly by the dynamic phase control.

Acknowledgements

This work was partly supported by MEXT, JSPS, ILE/Osaka University, CORE/Utsunomiya University, and U. S.-Japan Fusion Research Collaboration Program conducted by MEXT, Japan. This work was also partially supported by the project HiFi (CZ.02.1.01/0.0/0.0/15_003/0000449) and project ELI: Extreme Light Infrastructure (CZ.02.1.01/0.0/0.0/15_008/0000162) from European Regional Development. Computational resources were partially provided by the ECLIPSE cluster of ELI Beamlines.